\documentstyle[amsfonts,amssymb,amsmath,amsthm,prb,aps,12pt]{revtex}

\makeatletter

\def\@pnumwidth{2em}

\makeatother

\newtheorem{thm}{Theorem}
\newtheorem{cor}[thm]{Corollary}
\newtheorem{lem}[thm]{Lemma}

\theoremstyle{definition}
\newtheorem{defn}[thm]{Definition}
\newtheorem{ex}[thm]{Example}
\theoremstyle{remark}
\newtheorem{rem}[thm]{Remark}
\newcommand{\Lie}[2]{\underset{{#1}}{{\mathcal{L}}} {{#2}}}
\newcommand{\D}{{\rm D}}
\newcommand{\ud}[3]{{#1}_{\,\,\, {#3}}^{{#2}}}

\renewcommand{\d}[2]{{#1}_{\mathrm{#2}}}
\renewcommand{\u}[2]{{#1}^{\mathrm{#2}}}

\renewcommand{\baselinestretch}{1.37}

\begin{document}

\title{Multi--Lagrangians, Hereditary Operators and Lax Pairs for the Korteweg--de Vries Positive and Negative Hierarchies}%
\author{Miguel D. Bustamante${}^{{a)}}$\footnote[0]{E-mail: miguelb@macul.ciencias.uchile.cl\,,$\quad$ shojman@creavirtual.org
} and Sergio A. Hojman${}^{{b),\,a),\,c)}}$}
\address{${}^{{a)}}$ Departamento de F\'{\i}sica, Facultad
de Ciencias, Universidad de Chile,\\ Casilla 653, Santiago, Chile}
\address{ ${}^{{b)}}$ Centro de Recursos Educativos
Avanzados, CREA,\\ Vicente P\'erez Rosales 1356-A, Santiago,
Chile}
\address{ ${}^{{c)}}$ Facultad de Educaci\'on,
Universidad Nacional Andr\'es Bello,\\ Fern\'andez Concha 700,
Santiago, Chile}
\date{7 July 2003}
\maketitle
% AQUÍ VA LA COSA QUE SE PEGARÁ
%(Received
\begin{abstract}
We present an approach to the construction of action principles
(the inverse problem of the Calculus of Variations), for first
order (in time derivatives) differential equations, and generalize
it to field theory in order to construct systematically, for
integrable equations which are based on the existence of a
Nijenhuis (or hereditary) operator, a (multi--Lagrangian) ladder
of action principles which is complementary to the well--known
multi--Hamiltonian formulation. We work out results for the
Korteweg--de Vries (KdV) equation, which is a member of the
positive hierarchy related to a hereditary operator. Three
negative hierarchies of (negative) evolution equations are defined
naturally from the hereditary operator as well, in a concise way,
suitable for field theory. The Euler--Lagrange equations arising
from the action principles are equivalent to deformations of the
original evolution equation, and the deformations are obtained
explicitly in terms of the positive and negative evolution
vectors. We recognize, after appropriate coordinate
transformations, the Liouville, Sinh--Gordon, Hunter--Zheng and
Camassa--Holm equations as negative evolution equations. The
multi--Lagrangian ladder for KdV is directly mappable to a ladder
for any of these negative equations and other positive evolution
equations (e.g., the Harry--Dym and a special case of the
Krichever--Novikov equations). For example, several nonequivalent,
nonlocal time--reparametrization invariant action principles for
KdV are constructed, and a new nonlocal action principle for the
deformed system Sinh--Gordon $+$ spatial translation vector is
presented. Local and nonlocal Hamiltonian operators, are obtained
in factorized form as the inverses of all the nonequivalent
symplectic $2$--forms in the ladder. Alternative Lax pairs for all
negative evolution vectors are constructed, using the negative
vectors and the hereditary operator as only input. This result
leads us to conclude that, basically, all positive and negative
evolution equations in the hierarchies share the same
infinite--dimensional sets of local and nonlocal constants of the
motion for KdV, which are explicitly obtained using symmetries and
the local and nonlocal action principles for KdV.

\end{abstract}
{~~}\\


\noindent \textbf{Pacs Numbers}:
02.30.Zz,\,02.30.Ik,\,11.30.-j,\,11.10.Lm\,.

\newpage

\section{Introduction}
Hereditary or Nijenhuis operators \cite{Fuc79,And82,Dor93} play an
important role in the description of integrable systems: in terms
of these operators, the very definition of the positive and
negative hierarchies of integrable evolution equations may be
given, \cite{Fuc80} and they are used to construct, for these
equations, symmetries, \cite{Fuc80,Boi84} constants of the motion,
\cite{Boi84} alternative Lax pairs, \cite{{Bru95}}
multi--Hamiltonian structures \cite{Mag77} and weakly--nonlocal
multi--symplectic and multi--Hamiltonian structures.
\cite{Enr93,Mal01a,Mal01b}

The problem of constructing multi--Lagrangian structures (i.e., an
infinite ladder of action principles) for the Korteweg--de Vries
(KdV) equation has been tackled recently \cite{Nut00,{Nut01}} in
the context of localizable multi--Lagrangian structures, using the
bi--Hamiltonian formulation. However, the explicit expression of
the action principles associated to each symplectic $2$--form
requires the integration of the respective $2$--forms, a task
which is increasingly difficult as we move to the positive end of
the ladder, because of the increasing complexity of the
differential terms of the (weakly--nonlocal \cite{Mal01b})
$2$--forms. On the other hand, the symplectic $2$--forms are
increasingly nonlocal as we move to the negative end of the
ladder. The known way \cite{Nut01,Pav95} to get rid of the
nonlocality problem is to write the action principles in a
``local" coordinate system (Darboux coordinates) depending on the
specific symplectic $2$--form in the ladder, a process that gets
recurrently harder as we move to the negative end. Then, again,
the $2$--form must be integrated by hand in order to get the
action principle.

In this work, we make use of the Galilean symmetry \cite{Sch96}
and the factorized form of the hereditary operator \cite{Gut93}
for the KdV equation, to construct explicitly the action
principles for KdV in the positive and negative parts of the
ladder. No integration of any $2$--form is needed, nor is the
search for a special coordinate system. The factorized form of the
symplectic $2$--forms allows for the interpretation of the
resulting Euler--Lagrange equations (arising from each action
principle) as deformed equations, with flows given by KdV $+$
vectors in the positive and negative hierarchies, which are
computed explicitly.

Explicit expressions for local and nonlocal constants of the
motion for KdV are obtained using symmetries along with the local
and nonlocal action principles.

From the action principles obtained for the KdV equation we
construct action principles for flows defined by other positive
and negative vectors. In particular, a new nonlocal action
principle for the Sinh--Gordon (ShG) equation \cite{Rub70} (a
negative equation) are constructed.

It is a known result \cite{Bru95} that alternative Lax pairs for
the KdV equation and for positive KdV flows may be constructed
from the hereditary operator. Here we do the same construction for
all the negative KdV flows, and we conclude that the local and
nonlocal constants of the motion for KdV, define conserved
currents and constants of the motion for all the negative flows as
well.

The results here may be mapped to the following equations: a
special case of Krichever--Novikov, \cite{Sok84,Mal01a}
Harry--Dym, \cite{Gol91} Camassa--Holm, \cite{Hon99}
Hunter--Zheng, \cite{Bru02} ShG \cite{Rub70} and Liouville, all of
which are essentially flows belonging to the KdV positive or
negative hierarchies. We stress that the results are quite general
and may be extended to other systems related to hereditary
operators (e.g., nonlinear Schr\"odinger equation).

This paper is organized as follows: section \ref{sec:Pre} presents
a preview and notation for the method of construction of action
principles for given differential evolution equations, and a brief
survey of symmetries and constants of the motion in this context.
Next, we show the relationship of these Principles with
Hamiltonian theories, and finally we introduce the hereditary
property with the consequent construction of the positive and
negative hierarchies of integrable evolution equations. In section
\ref{sec:Ladders} we present and prove theorems on the explicit
construction of ladders of action principles and constants of the
motion, based on the hereditary operator and the Galilean symmetry
(whose definition is quite general and not restricted to KdV), and
we show how these action principles give rise to Euler--Lagrange
equations which are deformations of the original equations, due to
the fact that the symplectic operators have a nonzero kernel.

Section \ref{sec:example KdV} is devoted to examples of the above
constructions for the KdV equation. We obtain concise expressions
for the negative vectors, for the action principles, and for the
deformed Euler--Lagrange equations. Symplectic operators are
presented in factorized form (this allows for factorized
expressions for the Hamiltonian operators). Some known integrable
evolution equations are identified within the negative
hierarchies. Nonlocal constants of the motion for KdV are
concisely obtained using the internal symmetries and the negative
action principles. In section \ref{sec:results negative} we work
out examples of new nonlocal action principles for the ShG
equation, which in this context is identified as a negative
vector; we construct Lax pairs for the negative equations, thus
showing that the local and nonlocal constants of the motion for
KdV also work for negative equations. Finally, some
concluding remarks are presented in section \ref{sec:conc}.\\

For simplicity, we work in finite--dimensional notation. All
assertions and theorems in sections \ref{sec:Pre} and
\ref{sec:Ladders} are valid in finite dimensions, and they can be
extended to the case of field theory in all the instances dealt
with in this paper. We have used a Mathematica code of our
invention, to confirm the validity of some of the obtained local
and nonlocal action principles, symmetries and constants of the
motion for KdV and related equations.

\section{Preview and Notation}
\label{sec:Pre} Consider the autonomous equations of motion
\begin{equation}\label{eq:motion} \dot{q}^{\textrm{a}}(t) =
V^{\textrm{a}}[q^{\textrm{b}}(t)]\,,\quad{\textrm{a}} \in A\,.
\end{equation}
$A$ is a given ordered set called ``label set": the elements of it
label the degrees of freedom of the theory. From now on, we
suppress the dependence of the coordinates
$\{q^{{\textrm{a}}}\}_{{\textrm{a}} \in  A}$ on time when it is
obvious.

\begin{ex} The KdV Equation for the field $u(x,t)$, \mbox{$x \in [x_-,x_+]$,}
$t\in {\mathbb{R}}$, is
\begin{equation}\label{eq:KdV}
u_t = - u_{xxx} -12\, u\, u_x\,
\end{equation}
(suffixes denote partial differentiation). The label set is
${\mathrm{A}} = [x_-,x_+]$, and $x\in \mathrm{A}$ is a continuous
index.

We will use standard boundary conditions for the field:
$u,\,u_x,\,\ldots \to 0\,$ as $x \to x_\pm$, and we will set
$x_\pm = \pm \infty$, although the methods may be extended for the
treatment of other boundary conditions as well (in which case the
Weiss action principle \cite{Sud74} and the Witten--Zuckerman
$2$--form \cite{Wit86} come into play).
\end{ex}

The evolution equation (\ref{eq:motion}) is naturally defined on a
vector space spanned by the derivatives
$\{\frac{\partial}{\partial \u{q}{a}}\}_{{\mathrm{a}} \in
{\mathrm{A}}}$ (for the infinite--dimensional case, partial
derivatives with respect to the coordinates become functional
derivatives). We call $V = V\u{}{a}\,\frac{\partial}{\partial
\u{q}{a}}$ the \textbf{flow} vector or evolution vector for the
system (\ref{eq:motion}), where here, and throughout in this
paper, Einstein summation convention over repeated indices is
assumed (for the infinite--dimensional case, the summation is
extended to an integration over continuous indices).

\subsection{Action Principles}

The equations of motion (\ref{eq:motion}) are related to a
\textbf{Variational Principle} with \textbf{Action}
\begin{equation} \label{eq:action}
 {S}[q\u{}{a}(t),t] =
\int_{t_-}^{t_+} dt\,\bigg(P\d{}{a}\,\left( \dot{q}\u{}{a} -
V\u{}{a} \right)+ K\bigg)\,,
\end{equation}
 where the $1$--form $P[q\u{}{b}]$ and the zero-form $K[q\u{}{b}]$
satisfy the following equation:
\begin{equation*}
   P\d{}{a,b} V\u{}{b} + P\d{}{b} V\u{}{b}{}\d{}{, a}  =
 K\d{}{ , a}\,,
\end{equation*}
with $K\d{}{ , a}\equiv \frac{\partial K}{\partial q\u{}{a}}\,.$

We rewrite the above equation in terms of invariant structures:
\begin{equation}\label{eq:pair}
 \Lie{V}{} P = \delta K\,,
\end{equation}
where $\Lie{V}{}$ is the \textbf{Lie derivative} along the vector
$V$, and $\delta$ is the \textbf{exterior differential} (see
\cite{Nak92} for a definition of these operators).

\begin{defn}
We call the pair $(P;\,K)$ a \textbf{standard Lagrangian pair} for
$V$ if $K \neq 0$. In the special case $K = 0$ we call $P$ a
\textbf{non--standard Lagrangian} ($1$--form) for $V$: the latter
case allows for the construction of constants of the motion in a
direct way \cite{Bus02} (see theorem \ref{thm:3}).
\end{defn}

\begin{rem}The above objects should not be confused with the usual
``Lagrangian density" ${{L}}[q,\,\dot{q},\,t] = P\d{}{a}\,\left(
\dot{q}\u{}{a} - V\u{}{a} \right)+ K\,,$ which is the thing that
is integrated in time to give the action: $S = \int {{L}}\,dt\,.$
The $1$--form $P$ is also understood as a momentum map.
\cite{Nut01} When $P$ is a non--standard Lagrangian, it solves the
equation for a ``conserved covariant." \cite{Boi84}
\end{rem}

The general case of objects which depend explicitly on time is
easily worked out, \cite{Hoj81} but there is no need to do so in
the applications of this paper. Nevertheless, for symmetries and
constants of the motion the explicit time dependence will be
necessarily taken into account. In the sequel, we give the name
\textbf{time--(in)dependent} to those objects which do (not)
depend \textbf{explicitly} on time.

The \textbf{Euler--Lagrange} equations which come from the action
(\ref{eq:action}) are:
\begin{equation*}
\Sigma\d{}{a b} \left( \dot{q}\u{}{b} - V\u{}{b}\right) = 0\,,
\end{equation*}
where $\Sigma \equiv \delta P$ is the \textbf{symplectic $2$--form
}or Lagrange bracket whose components are:
\begin{equation*}
\Sigma\d{}{a b} =  P\d{}{b,a}- P\d{}{a,b}\,.
\end{equation*}
It is worth mentioning that, in \cite{Wit86}, a symplectic
$2$--form is induced by an action principle in essentially the
same way we have derived the above symplectic $2$--form from the
action principle (\ref{eq:action}).

Notice that these Euler--Lagrange equations do not imply the
original equations of motion (\ref{eq:motion}); instead they imply
\textbf{deformed} or mixed equations, where the deformation is
represented by an additive extra term which is an arbitrary linear
combination of vectors belonging to the \textbf{kernel} of the
symplectic $2$--form. In the case of KdV, we will obtain the
deformations explicitly. See \cite{Bus02} for examples in the
finite dimensional case.

The symplectic $2$--form associated to this action principle is
easily seen to satisfy
\begin{eqnarray}
\nonumber \delta\, \Sigma &=& 0 {\textrm{\quad (closure)}}\\
\label{eq:symp2} \Lie{V}{\,} \Sigma &=& 0\,,
\end{eqnarray}
therefore the inverse process could be done: starting from a
symplectic $2$--form $\Sigma$ for the flow vector $V$, we
construct the standard Lagrangian $1$--form, from $\delta\,P =
\Sigma$ and the $0$--form $K$ is obtained by integration of
equation (\ref{eq:pair}). This process suffers from technical
difficulties, which increase when the objects are infinite
dimensional and nonlocal. Fortunately, for the KdV equation there
is a constructive way of finding the action principles (see
theorems \ref{thm:1} and \ref{thm:2}).

\subsection{Hamiltonian theories are induced from Symplectic Structures}

The relationship of the symplectic $2$--form with the Hamiltonian
formulation is very simple: \cite{Olv93} consider the formal
inverse (i.e., except for a finite kernel that the operators may
possess) of the above $2$--form, the time--independent $(2,0)$
tensor $J$ such that $J\cdot \Sigma = \mathbb{I}\,.$ It is
possible to show that $\Sigma$ is closed if and only if $J$
satisfies the \textbf{Jacobi identity}, which we write in the
form:
\begin{equation}\label{eq:Jac}
  \Lie{J\cdot U}{J} = J\cdot \delta U\cdot J\,,\quad \forall \,\,1\textrm{--form }\,U\,.
\end{equation}
$J$ is known as a Hamiltonian operator or Poisson bracket. Now,
equation (\ref{eq:symp2}) implies $\Lie{V}{\,J} = 0\,.$ Therefore,
according to the Jacobi Identity (\ref{eq:Jac}), a Hamiltonian
theory for the flow $V$ is induced by the symplectic $2$--form
$\Sigma$: the equation $V=J\cdot U$ implies $\delta U=0$, thus $V
= J\cdot \delta H\,,$ where $H$ is the Hamiltonian, a
time--independent $0$--form.

\subsection{Symmetries}

Symmetries play a crucial role in the construction of the action
principles. A symmetry for the system (\ref{eq:motion}) is known
as a vector with components $\u{\eta}{a}$ that takes solutions
into solutions of equation (\ref{eq:motion}), in the sense that
given any solution ${q}^{\textrm{a}}(t)$ such that
$\dot{q}^{\textrm{a}}(t) = V^{\textrm{a}}[q^{\textrm{b}}(t)]\,,$
then $\u{\tilde{q}}{a} \equiv \u{q}{a} +
\epsilon\,\u{\eta}{a}[\u{q}{b},t]$ is also a solution up to order
$\epsilon^2$, i.e.: $\dot{\tilde{q}}{}^{\textrm{a}}(t) =
V^{\textrm{a}}[\tilde{q}^{\textrm{b}}(t)]\,+\,O(\epsilon^2)\,.$

It is easily seen \cite{Hoj96} that this condition leads to the
equation $ \frac{\partial}{\partial t}\u{\eta}{a}
 + \ud{\eta}{\mathrm{a}}{, \mathrm{b}}\,\u{V}{b} - \ud{V}{\mathrm{a}}{, \mathrm{b}}\,\u{\eta}{b} = 0
\,$ or, in a covariant way, $\left(\frac{\partial}{\partial t} +
\Lie{V}{}\right)\eta = 0\,.$

\begin{ex}
The Galilean and the dilatation symmetries
 for the KdV equation are defined respectively by
\begin{eqnarray}
\u{\d{\eta}{G}[u,t]}{x} &=& \frac{1}{8} - \frac{3}{2} \,t \,u_x\,, \label{eq:gal} \\
\nonumber \u{\d{\eta}{D}[u,t]}{x} &=& u + \frac{1}{2}\,x\,u_x - t
\left(\frac{3}{2}\,u_{xxx} + 18\,u\,u_x\right)\,.\
\end{eqnarray}
In \cite{Sch96}, there is an open question concerning the role of
the Galilean and dilatation symmetries in the construction of
constants of the motion for KdV. An answer to this question is
given in this work: these symmetries actually lead to action
principles for the KdV equation, which are involved in Noetherian
and non--Noetherian constructions \cite{Hoj96} of constants of the
motion (see  theorems \ref{thm:1}, \ref{thm:2}, and \ref{thm:3}).
\end{ex}

\subsection{Constants of the Motion}

A constant of the motion for the system (\ref{eq:motion}) is a
functional ($0$--form) $C[\u{q}{a},t]$ which is conserved in time
under the evolutionary system: $\left.\frac{\D}{\D
t}C[\u{q}{b},t]\right|_{\mathrm{on-shell}} \equiv
\frac{\partial}{\partial t}C + \d{C}{, a}\,\u{V}{a} = 0\,,$ where
the partial time derivative accounts for the explicit time
dependence and $\frac{\D}{\D t}$ denotes the convective or total
derivative along the variable $t$.

This equation is best written in a covariant way: \cite{Hoj96}
$\bigg( \frac{\partial}{\partial t} + \Lie{{V}}{} \bigg) C = 0\,.$

We will usually work with time--independent constants of the
motion: $\frac{\partial}{\partial t} C = \Lie{{V}}{\,} C  =  0\,.$

\subsection{The Hereditary Property: Hierarchies of Evolution Equations}

Many integrable systems are related to a Nijenhuis or hereditary
operator, which is a time--independent $(1,1)$ tensor $R$ that
solves: \cite{Olv93} $\Lie{R \cdot \eta}{R} = R \cdot
\Lie{\eta}{R}\,,\quad \forall \textrm{ vector } \eta\,.$

Out of the kernel of this operator, and of its inverse,
hierarchies of integrable evolution equations arise which are
symmetries of each other \cite{Fuc80} (this will be worked out in
detail for the KdV case later on).

According to \cite{Fuc80}, given a hereditary operator $R$ and a
flow vector field (labelled with a number) $V\d{}{1}$ such that
$\Lie{V\d{}{1}}{R}= 0\,,$ i.e., $R$ is a \textbf{recursion}
operator  for $V_1$, then a \textbf{hierarchy} is defined as a
semi--infinite collection of evolution vectors: $\{V_{j} =
R^{j-1}\cdot V_1 \,,\quad {j}=1,\,\ldots,\,\infty\}\,,$ which are
symmetries of each other: $ \Lie{V_{i}}{V_{j}} = 0\,,\quad
i,\,j\geq 1$, and thus every evolution vector in the hierarchy
defines an evolution equation which is integrable. In the KdV
hierarchy, the KdV equation is the second member ($V_2$). The
first vector ($V_1$) represents the translation symmetry, and it
is shown to generate the kernel of the inverse hereditary
operator, $R^{-1}$. By convention, we refer to the above as a
\textbf{positive hierarchy}.

The hereditary property for the operator $R\,$ may be used to show
formally that $R^{-1}$ is also hereditary. \cite{Gut93} Therefore,
we could conjecture that new hierarchies (referred to as
\textbf{negative hierarchies}) of evolution vectors may be
constructed, which first members generate the kernel of the
operator $R$, and successive members are defined by contraction of
the first members with powers of the operator $R^{-1}$. In the KdV
case, there are three negative hierarchies. These new negative
equations include the ShG, Liouville, Camassa--Holm, and the
Hunter--Zheng equations.

\subsection{Notation: the Positive and Negative Hierarchies in terms of the Hereditary Operators}

The analysis is restricted to the KdV hierarchies, but it is
easily generalizable to other systems related to hereditary
operators.

We will adopt the following notation for evolution vectors in the
hierarchies:
\[u_t = V\ud{}{(k)}{n}[u]\,,\]
where ${{k}}=1$ denotes the positive hierarchy; ${{k}} =
\,-1,\,-2,\,-3$ for the three negative hierarchies, and ${{n}} =
1,\ldots,\,\infty$ denotes the place of a vector within the
hierarchy, so that we have $n=1$ for the first vector of each
hierarchy, i.e., the relevant generator of the kernel of
$R^{-{\mathrm{sgn}}({k})}$:
\begin{equation}\label{eq:kerR}
\begin{array}{rcl}
R^{-1}[u] \cdot V\ud{}{(1)}{1}[u]& =& 0\,, \\
R[u] \cdot V\ud{}{(-1)}{1}[u] = R[u] \cdot V\ud{}{(-2)}{1}[u] &= &
R[u] \cdot V\ud{}{(-3)}{1}[u] =0\,. \
\end{array}
\end{equation}
Successive members in the hierarchies are defined by recurrence:
\[V\ud{}{(k)}{n+1}[u] = (R[u])^{{\mathrm{sgn}}({{k}})}\cdot V\ud{}{(k)}{n}[u]\,,\quad {{n}} \geq 1\,,\quad {{k} = 1,\, -1,\, -2,\, -3}\,.\]

In this way, the positive hierarchy begins with the vector
$V\ud{}{(1)}{1}[u] = -u_x$, continues with the KdV vector
$V\ud{}{(1)}{2}[u] = -u_{xxx}-12\,u\,u_x$, and so on (these
vectors were called $V_1$ and $V_{2}$ in the last subsection).

For the negative hierarchy, as the operator $R^{-1}$ is harder to
work with, there is a recurrent way of writing the negative
vectors, in terms of $R$:
\begin{equation}\label{eq:recNeg}
V\ud{}{(k)}{n}[u] = R[u] \cdot V\ud{}{(k)}{n+1}[u]\,,\quad {{n}}
\geq 1\,,\quad {{k} = -1,\,-2,\,-3}\,.
\end{equation}

The explicit expression for negative vectors relies on the
factorized form \cite{Gut93,Vil01} of the hereditary operator, and
will be realized in section \ref{sec:example KdV} in terms of
nonlocal fields which, however, are tractable in the same scheme
as the local ones.

\section{Ladders of Action Principles and Constants of the Motion}
\label{sec:Ladders} Complementary to the well--known
bi--Hamiltonian formulation, \cite{Fuc80} we may find a
bi--symplectic or multi--symplectic structure starting from the
hereditary property. Assume that we have a Nijenhuis operator $R$
along with one closed $2$--form $\Sigma^{(1)}$ such that
$\Sigma^{(2)} \equiv \Sigma^{(1)}\cdot R$ be a closed $2$--form:
then, the two semi--infinite dimensional sets (\textbf{symplectic
ladders}) of $2$--forms
\begin{equation*}
\{\Sigma^{(n)}\equiv \Sigma^{(1)}\cdot R^{{n-1}}\,,\quad
{n}=1,\,\ldots,\,\infty\} \quad {\textrm{(positive symplectic
ladder)}}\,,
\end{equation*}
and
\begin{equation*}
\{\Sigma^{(n)}\equiv \Sigma^{(1)}\cdot R^{{n-1}}\,,\quad
{n}=0,\,-1,\,\ldots,\,-\infty\} \quad {\textrm{(negative
symplectic ladder)}}\,,
\end{equation*}
contain only closed $2$--forms. The distinction between positive
and negative ladders is somewhat arbitrary, for it depends on
which hereditary operator, $R$ or $R^{-1}$, is being used, and
which symplectic $2$--form is taken as $\Sigma^{(1)}\,.$

The proof of the above statement is very simple. In fact, it is
equivalent to the proof for the so--called Poisson pencil or set
of compatible implectic operators \cite{Fuc80} for
multi--Hamiltonian theories, after defining the implectic or
Hamiltonian operators as the inverses of the $2$--forms in the
ladder.

The above result is independent of any evolution vector. When we
consider the vectors in the hierarchies, however, it is easily
checked (as it holds in the examples) that
$\Lie{V\d{}{1}}{\Sigma^{(1)}} = 0$. Therefore, using Leibnitz
rule, all the $2$--forms in the ladder are symplectic operators
for the first evolution vector in the hierarchy.

Using the identity $\Lie{R\cdot
\eta}{\Sigma}-\Lie{\eta}{\left(\Sigma \cdot R\right)} = i_{R
\cdot\eta}\delta\Sigma - i_{\eta}\delta(\Sigma \cdot R)\,,$ which
holds for any vector $\eta$, $(1,1)$ tensor $R$ and $2$--form
$\Sigma$, where $i_{\eta}$ stands for interior product
(contraction of the vector $\eta$ with the left component of a
$p$--form), we obtain the important result for any hierarchy:
\begin{equation}\label{eq:result-symplectic}
  \Lie{V_{j}}{\,\Sigma^{(n)}} = 0\,,\quad
  {j}=1,\,2,\,\ldots,\,\infty,\quad
  {n}=-\infty,\,\ldots,\,\infty\,,
\end{equation}
which means that a ladder of action principles may be constructed
for every evolution vector in the hierarchy (in particular, for
the KdV equation).

This fact is used in \cite{Nut01} to construct action principles,
with the only drawback it needs to integrate the $2$--forms
(Poincar\'e lemma) in order to get the action principles.

Let us assume for the rest of this section that we have a
Nijenhuis operator $R$ along with its inverse $R^{-1}$, a
symplectic ladder $\{\Sigma^{(n)}\,,\quad
{n}=-\infty,\,\ldots,\,\infty\}\,,$ and a hierarchy
$\{V_{j}\,,\quad {j}=1,\,\ldots,\,\infty\}\,$ with the
corresponding properties mentioned above.

The purpose of the following subsection is to construct, for the
second evolution vector in the positive hierarchy (though the
analysis is easily extended for other positive and negative
evolution vectors), the action principles associated to each of
the above symplectic $2$--forms. These action principles are
involved in the explicit construction of constants of the motion
for the evolution equation.

\subsection{Construction of Action Principles and Constants of the Motion out of Symmetries and Symplectic Operators}

Heuristically, if we had a symmetry for a given evolution equation
we could obtain in a direct way (by contraction of it with any
symplectic $2$--form) a Lagrangian $1$--form and therefore an
action principle for that equation.

We have done this procedure for any equation in the positive
hierarchy, using the Galilean symmetry, obtaining as a result a
ladder of action principles which are (explicitly) time--dependent
(in fact, linear in time). For simplicity, however, we rewrite the
actions as time--independent objects, and the discussion will be
restricted to the second vector (which corresponds to the KdV
equation), which is from now on referred to as the vector $V_2\,.$

The following definition will be a key to the construction of the
ladder of action principles for the evolution vector $V_2$, and it
permits a generalization to the negative hierarchies as well as to
other systems (e.g., nonlinear Schr\"odinger equation):

\begin{defn} The
Galilean vector field $\eta\d{}{gal}$ is a time--independent
vector field, defined by three properties:
\begin{equation}\label{eq:defn:gal}
  \begin{array}{rcl}
\Lie{\eta\d{}{gal}}{R} &=& \mathbb{I}\,,\\
\Lie{\eta\d{}{gal}}{\Sigma^{(1)}} &=& 0\,,\\
\Lie{\eta\d{}{gal}}{V_{2}}&=&\alpha\,V_1\,,
  \end{array}
\end{equation}
where $\alpha$ is a numeric constant.
\end{defn}
As a consequence of the definition, it turns out that
$\eta\d{}{gal}$ is a Mastersymmetry \cite{Oev89} for the hierarchy
$\{V_{j}\,,\quad {j}=1,\,\ldots,\,\infty\}\,.$ Explicitly, we have
$ \Lie{V_{j+1}}{\eta\d{}{gal}} = (\alpha + {j} -1)\,V_{j}\,,$ for
$ {j}=1,\,\ldots,\,\infty\,.$

The aim is to construct time--independent standard Lagrangian
pairs $(P;\,K)$, where $ \Lie{V_2}{P} = \delta K\,.$ The action
principles will read $S[\u{q}{a}(t)] = \int_{t_-}^{t_+} dt\,
\left(\d{P}{a} \left(\dot{q}^{{\mathrm{a}}} -
V_2^{{\mathrm{a}}}\right)+ K\right)\,, $ and the Euler--Lagrange
equations will involve the symplectic $2$--forms in the ladder,
i.e., $\delta P = \Sigma\,.$

\begin{thm}\label{thm:1}
The $1$--forms defined by $P^{(m)} \equiv
i_{\eta\d{}{gal}}\,\Sigma^{(m+1)}\,,$ for $
  {m}=-\infty,\,\ldots,\,\infty\,$
are ``integrals" of the symplectic $2$--forms in the ladder. That
is to say,
\begin{equation}\label{eq:lagr-Lad}
  \delta P^{(m)}=  {m}\,\Sigma^{(m)}\,,\quad
  {m}=-\infty,\,\ldots,\,\infty\,.
\end{equation}
\end{thm}

\begin{proof}
If we take the exterior derivatives of the $1$--forms, using the
identity
\begin{equation}\label{eq:id-p-form}
  \Lie{\eta}{} = i_\eta\,\delta +\delta \,i_\eta\,,
\end{equation}
which holds (for every vector $\eta$) when operating on any
$p$--form, we find
\begin{equation*}
\delta P^{(m)} = \Lie{\eta\d{}{gal}}{}\Sigma^{(m+1)} =
\Lie{\eta\d{}{gal}}{}(\Sigma^{(1)}\cdot R^{m}) =
\Sigma^{(1)}\cdot \Lie{\eta\d{}{gal}}{}(R^{m}) = {m}\,\Sigma^{(1)}
\cdot R^{m-1} = {m}\,\Sigma^{(m)}  \,,{{m}} \in {\mathbb{Z}}
\end{equation*}
after using the definition of Galilean vector and Leibnitz rule.
\end{proof}

In order to complete the action principles, there remains to find
the second members of the corresponding standard Lagrangian pairs.

\begin{thm}\label{thm:2}
For each $m \in \mathbb{Z}$, the pair $(P^{(m)};K^{(m)})$ with
$K^{(m)} \equiv \frac{\alpha}{m + \alpha}\,i_{V_2}P^{(m)}\,,$ is a
standard Lagrangian pair for the evolution equation
$\dot{q}\u{}{a} = V_2\u{}{a}\,,$ i.e., $\Lie{V_2}{P^{(m)}} =
\delta K^{(m)}\,.$ For $m \neq 0$, the action principle is
\begin{equation}\label{eq:action-KdV}
  S^{(m)}[\u{q}{a}(t)] = \int_{t_-}^{t_+} \,\d{P^{(m)}}{a}\,\left(
\dot{q}^{{\mathrm{a}}}- \frac{m}{m +
\alpha}\,V_2\u{}{a}\right)\,dt\,.
\end{equation}

Moreover, the $0$--forms $K^{(m)}$ are constants of the motion for
the evolution equation, for $m \in \mathbb{Z}$:
\begin{equation*}
\Lie{V_2}{K^{(m)}} = 0\,.
\end{equation*}

\end{thm}

The above is a Noetherian way to construct constants of the
motion, for all the action principles in the ladder for KdV are
naturally invariant under the KdV flow $V_2$ itself (see equation
(\ref{eq:result-symplectic})).

\begin{rem}The case ${m}=0$ would lead to a trivial action principle from
equation (\ref{eq:action-KdV}), for the Euler--Lagrange equations
are identically zero: it is shown that this case leads to a
time--dependent constant of the motion. This does not mean that
the symplectic $2$--form $\Sigma^{(0)}$ defines a trivial action
principle. In fact, its associated action principle may be found
by hand (see the end of this subsection), and it is related to the
usual action principle for the ShG equation (see subsection
\ref{subsec:action-sinhg}).
\end{rem}

\begin{proof}[Proof: Lagrangian Pairs]
For $m \in \mathbb{Z}$, take Lie derivatives of the $1$--forms
$P^{(m)}$ along the evolution vector $V_2$, using Leibnitz rule:
\begin{equation}\label{eq:Lagr-1-KdV}
  \Lie{V_{2}}{P^{(m)}} = \Lie{V_{2}}{}i_{\eta\d{}{gal}} \Sigma^{(m+1)}
  = -\alpha\, i_{V_1}\Sigma^{(m+1)} = -\alpha \,i_{V_{2}}\Sigma^{(m)}\,.
\end{equation}
But, using the identity (\ref{eq:id-p-form}) and the result
(\ref{eq:lagr-Lad}) we rewrite the last expression to get
\begin{equation*}
m \,\Lie{V_2}{P^{(m)}} =-\alpha \left(\Lie{V_{2}}{P^{(m)}} -
\delta\,
  i_{V_{2}}P^{(m)}\right)\,,
\end{equation*}
therefore
\begin{equation}\label{eq:Lagr-pair-KdV}
  \Lie{V_{2}}{}P^{(m)} = \delta \left(\frac{\alpha}{m +
\alpha}\,i_{V_{2}}P^{(m)}\right) \equiv \delta  K^{(m)}\,.
\end{equation}
\end{proof}

\begin{proof}[Proof: Constants of the Motion]
We use the above result (\ref{eq:Lagr-pair-KdV}), to find
\begin{equation*}
\Lie{V_{2}}{K^{(m)}} = \frac{\alpha}{m +
\alpha}\,i_{V_{2}}\Lie{V_{2}}{P^{(m)}} = \frac{\alpha}{m +
\alpha}\, i_{V_{2}}\delta K^{(m)} = \frac{\alpha}{m + \alpha}\,
\Lie{V_{2}}{} K^{(m)}\,,
\end{equation*}
which implies $\Lie{V_{2}}{K^{(m)}} = 0\,,\quad {m} \neq 0\,.$

For $m =0\,$ we find a weaker result: Equation
(\ref{eq:Lagr-1-KdV}) implies $\delta \Lie{V_{2}}{K^{(0)}} = 0\,.$
Therefore $\Lie{V_{2}}{K^{(0)}} = c\,,$ is a number (usually equal
to zero) that may be absorbed to define a time--dependent constant
of the motion: $\tilde{K}^{(0)}(t) = {K}^{(0)} - c\,t\,.$
\end{proof}

For the KdV equation, when ${m}\geq -1$ we get the usual
denumerably infinite set of constants of the motion. \cite{Eil81}
Notice that this theorem represents also a constructive method to
obtain such constants. On the other hand, when $m\leq -2$ the
constants are numerical or vanishing boundary terms. Amazingly,
this fact allows one to construct an infinite number of nonlocal
constants of the motion for KdV, using the nonlocal action
principles (see subsection \ref{subsec:nonlocal-const-KdV}).

\begin{proof}[Proof: Action Principles]
The action principles (\ref{eq:action-KdV}) arise directly from
equation (\ref{eq:action}), using the definition of $K^{(m)}$.
\end{proof}

From the point of view of theorem \ref{thm:2}, the case ${m}=0$
also leads to a time--dependent constant of the motion. From
equation (\ref{eq:lagr-Lad}), it follows that $P^{(0)} = \delta
C^{(0)}\,,$ and thus \linebreak $\Lie{V_{2}}{}C^{(0)} =
K^{(0)}\,.$ But we know that $\Lie{V_{2}}{}K^{(0)} = c\,$ is a
number. We obtain the following time--dependent constant of the
motion for the evolution vector $V_{2}$:
\begin{equation}\label{eq:t--d const}
  C[q\u{}{a}(t),\,t] = C^{(0)} - t\,K^{(0)}+ \frac{c\,t^2}{2}\,.
\end{equation}

Finally, for the case ${m}=0\,,$ a special (``missing") action
principle is constructed by hand from integration of the $2$--form
$\Sigma^{(0)}\,,$ which leads to the $1$--form $P^{(M)}\,,$ such
that $\delta P^{(M)} = \Sigma^{(0)}\,.$ We will have
$\Lie{V_{2}}{P^{(M)}} = \delta K^{(M)}\,,$ and the action is
\begin{equation*}
  S^{(0)}[q\u{}{a}(t)] = \int_{t_-}^{t_+} \, \left(\d{P^{(M)}}{a}\,\left(
\dot{q}^{{\mathrm{a}}}- V_{2}\u{}{a}\right) +
K^{(M)}\right)\,dt\,.
\end{equation*}

\subsection{The Euler--Lagrange Equations as Deformed Evolution Equations}

The Euler--Lagrange equations that arise from variation of each
action $S^{(m)}\,,\quad {m} \in \mathbb{Z}\,$ are, apart from
nonzero numeric factors,
\begin{equation*}
  \Sigma^{(m)}\d{}{a b}(\dot{q}\u{}{b} - V_{2}\u{}{b}) = 0\,,\quad {m} \in \mathbb{Z}\,.
\end{equation*}
The kernel of the symplectic operators, ${\mathrm{Ker}}
\Sigma^{(m)}\,,$ is of importance here. For each action principle
we obtain an equivalent, \textbf{deformed}, evolution equation:
\begin{equation*}
  \dot{q}\u{}{a} = V_{2}\u{}{a} + \sum_{j=1}^{N_{m}}
  \theta_j\,\eta_{j;{m}}\u{}{a}\,,
\end{equation*}
where $N_{m} = {\mathrm{dim}}({\mathrm{Ker}} \Sigma^{(m)})\,,$ the
vectors $\{\eta_{j;{m}}\}_{j=1}^{N_{m}}$ generate the kernel of
$\Sigma^{(m)}\,,$ and $\theta_j = \theta_j(t)$ are arbitrary
$0$--forms: it can be said that these Euler--Lagrange equations
and the action principles acquire extra symmetries (as compared to
the symmetries of the original equations).

As the $2$--forms here are formed by contraction of powers of the
hereditary operators $R$ and $R^{-1}$ with $\Sigma^{(1)}$, it is
clear that the kernel of the $2$--forms are computed essentially
from vectors in the kernel of the operators $R^m$ and
$R^{-m}\,,\,$ for $m>0:$ as we have mentioned, these are the
positive and negative evolution vectors. In section
\ref{sec:example KdV}, we will find explicitly the deformed
equations for the KdV equation in terms of the positive and
negative vectors.

\subsection{Construction of nonlocal Constants of the Motion from Symmetries and non--standard Lagrangian $1$--forms}
\label{subsec:nonlocal-const-KdV} Let us assume, as it will be
demonstrated in subsection \ref{subsec:neg lag KdV} for the KdV
case (under usual boundary conditions), that the constants of the
motion from theorem \ref{thm:2} are $K^{(m)} = 0$ or a numeric
constant for $m \leq -2\,.$ This implies that $\Lie{V_2}{}P^{(m)}
= 0\,,$ i.e., $P^{(m)}$ is a non--standard Lagrangian $1$--form
for the flow $V_2$. Assume also that the evolution equation
defined by the flow $V_2$ possesses a symmetry $\eta\,.$ Then
\begin{thm}\label{thm:3}The $0$--forms defined by $Q^{(m)} \equiv i_\eta P^{(m)}\,,$
for $m \leq -2\,,$ are constants of the motion for the flow
$V_2\,,$ i.e., $\left(\partial_t + \Lie{V_2}{}\right){Q^{(m)}} =
0\,$ for $m \leq -2\,.$
\end{thm}
\begin{proof}The proof follows directly from Leibnitz rule.
\end{proof}
The above is a non--Noetherian way \cite{Hoj96} to construct
constants of the motion, in the sense that the action principles
need not be invariant under the relevant symmetry. In subsection
\ref{subsec:nonlocal-const-KdV-2} we construct ``generating
functions" for three infinite--dimensional sets of nonlocal
constants of the motion for the KdV equation, setting $\eta$ as a
nonlocal internal symmetry for KdV.

\section{Example: The KdV Equation}
\label{sec:example KdV}
\subsection{Known Objects}
We begin by presenting the Nijenhuis operator relevant for the KdV
hierarchy: as an operator, \cite{Fuc80} $R[u] = \D^2 + 8\, u + 4\,
u_x \,\D^{-1}\,,$ where $\D$ and $\D^{-1}$ are, respectively, the
derivative and the anti--derivative operators: $\D f(x)  \equiv
\frac{\partial f(x)}{\partial x}\,,$ $\,\,\D^{-1}g(x)  \equiv
\int_{x_-}^{x_+}\epsilon(x-x')\,g(x')\,dx'\,,$ with
$\epsilon(x-x') = 1/2\,{\mathrm{sign}}(x-x')$.

Next, the  \textbf{positive} hierarchy begins with the vector
$V\ud{}{(1)}{1}[u] = -u_x\,.$ The second vector in the positive
hierarchy is obtained after application of the Nijenhuis operator
on the latter vector: $V\ud{}{(1)}{2}[u] = R[u]\cdot
V\ud{}{(1)}{1}[u] =
  -u_{xxx}-12\,u\,u_x \equiv V_2[u]\,.$ We see it represents the KdV equation (\ref{eq:KdV}).

Next, we write the first symplectic $2$--form: as an operator,
\cite{Fuc80} $\Sigma^{(1)}[u] = \D^{-1}\,.$ The second symplectic
operator is constructed just by contracting the latter operator
with the Nijenhuis operator: $ \Sigma^{(2)}[u] =   \Sigma^{(1)}[u]
\cdot R[u] = \D + 4\,u\,\D^{-1} + 4\,\D^{-1}\,u\,.\,$ These
operators are closed under usual boundary conditions for the
vector fields: the ladder, then, contains only closed $2$--forms.
Now, it is easy to show that these operators are symplectic for
the flow defined by $V\ud{}{(1)}{1}$, therefore all operators in
the ladder are symplectic for the KdV flow $V\ud{}{(1)}{2}$, as it
is stated in equation (\ref{eq:result-symplectic}).

Finally, the Galilean vector is just the time--independent part of
the Galilean symmetry (\ref{eq:gal}): $ \eta\d{}{gal}[u] =
\frac{1}{8}\,, $ and the constant in the last of the defining
equations (\ref{eq:defn:gal}) is $\alpha = 3/2\,.$

\subsection{Explicit form of the KdV Negative Hierarchies: Linear
generalization and factorization of the Hereditary Operator}

In order to find explicitly the KdV \textbf{negative} hierarchies,
we factorize a generalization of the hereditary operator $R$,
which is obtained by addition of a multiple of the identity tensor
${\mathbb{I}}$:
\begin{equation*}
  R(\lambda)[u] \equiv R[u] + 4\, \lambda\,{\mathbb{I}}\,.
\end{equation*}
This is also a hereditary operator, for fixed $\lambda$, which is
taken as an arbitrary real number.

The idea behind this generalization, is that the kernel of
$R(\lambda)[u]$ contains all the negative hierarchies in its
Taylor expansion around $\lambda= 0\,,$ so we will write the
negative hierarchies in a compact way.

\begin{lem}\label{lemLeft}
The vectors
$V\ud{}{(-1)}{1}(\lambda)[u],\,V\ud{}{(-2)}{1}(\lambda)[u],\,V\ud{}{(-3)}{1}(\lambda)[u]$,
defined by
\[V\ud{}{(k)}{1}(\lambda)[u] \equiv \sum_{{{n}}=0}^{\infty}(-4\,\lambda)^{{n}}
\,V\ud{}{(k)}{n+1}[u]\,,\quad {{k}} = -1,\,-2\,-3\,\]
 generate the vectorial kernel of $R(\lambda)[u]$.

Conversely, all vectors in the negative hierarchies may be
obtained from the vectorial kernel of $R(\lambda)[u]$:
\begin{equation}\label{eq:neg(-n) from neg(-1)}
V\ud{}{(k)}{n+1}[u] = \frac{1}{{n}!}(-4)^{-{n}}
\left.\frac{\partial^{{n}}}{\partial\lambda^{{n}}}
V\ud{}{(k)}{1}(\lambda)[u]\right|_{\lambda = 0}\,,\quad {{k}} =
-1,\,-2\,-3 \,,\quad {{n}} = 0,\,1,\ldots,\,\infty\,.
\end{equation}
\end{lem}

\begin{proof}Consider the action of $R[u]$ on the vector
$V\ud{}{(k)}{1}(\lambda)[u]\,.$ Using equations (\ref{eq:kerR})
and (\ref{eq:recNeg}), we get
\begin{equation*}
  R[u] \cdot V\ud{}{(k)}{1}(\lambda)[u] = \sum_{{{n}}=0}^{\infty}(-4\,\lambda)^{{n}}
\,  R[u] \cdot V\ud{}{(k)}{n+1}[u] =
\sum_{{{n}}=1}^{\infty}(-4\,\lambda)^{{n}} \, V\ud{}{(k)}{n}[u] =
-4 \,\lambda\, V\ud{}{(k)}{1}(\lambda)[u]\,,
\end{equation*}
therefore $R(\lambda)[u]\cdot V\ud{}{(k)}{1}(\lambda)[u] =
(R(\lambda)[u] + 4 \,\lambda )\cdot V\ud{}{(k)}{1}(\lambda)[u] =
0\,.$
\end{proof}

The factorization process \cite{Vil01} implies the definition of
auxiliary fields, which are directly related to nonlocal
pre--potentials found in the literature \cite{Gut93} and to the
associated isospectral linear eigenvalue problem: \cite{Lax68}
\begin{equation}\label{eq:lin-eig}
\psi_{xx} + 2 \, u \, \psi = -\lambda\,\psi\,,
\end{equation}
where $\psi = \psi(x,t;\lambda)\,.$ As usual, we assume $\lambda_t
= 0\,,$ and $u = u(x,t)$ is independent of $\lambda$.

Alternatively, we write the above equation as: $L(\lambda)\cdot
\psi = 0$, where $L(\lambda) = \D^2+2\, u +\lambda$ or, in a
factorized way, $ L(\lambda) =
\frac{1}{\psi}\,\D\,\psi^2\,\D\,\frac{1}{\psi}\,,$ is the
\textbf{Lax operator}. The elements in the kernel of this operator
are solutions of the linear problem (\ref{eq:lin-eig}). Two l.i.
solutions are: $\psi(\lambda) \equiv \psi(x,t;\lambda)\,,$ and
\begin{equation}\label{eq:2ndSol}
  \bar{\psi}(\lambda) \equiv {\psi}(\lambda)
\D^{-1}(1/{\psi}(\lambda)^2)\,.
\end{equation}
The above eigenvalue problem may be understood as an extended
coordinate system labelled by $\psi(x,t;\lambda)$, with $\lambda$
as an additional variable (just like $x$), and which however must
solve an extra equation, $L(\lambda)\cdot \psi(\lambda) = 0$,
which we call \textbf{constraint}. This constraint lets us write
derivatives of the field $\psi(\lambda)$ with respect to $\lambda$
in terms of the field itself, in a nonlocal way. This will be
useful in the next section, when we write the negative ladder of
action principles. We obtain, apart from integration constants,
\begin{equation}\label{eq:local--nonlocal}
\begin{array}{rcl}
\frac{\partial^n}{\partial \lambda^{n}} \psi
&=&(-1)^n\,n!\,L(\lambda)^{-n}\,\psi\,,\\
\frac{\partial^n}{\partial \lambda^{n}} \bar{\psi}
&=&(-1)^n\,n!\,L(\lambda)^{-n}\,\bar{\psi}\,,\quad n\geq 1\,,
\end{array}
\end{equation}
where $L(\lambda)^{-1} =
{\psi}\,\D^{-1}\,\frac{1}{\psi^2}\,\D^{-1}\,{\psi}\,,$  $\psi =
\psi(\lambda)$ and $\bar{\psi} = \bar{\psi}(\lambda)\,.$

Now, the factorization of $R(\lambda)[u]$ is found to be
\[R(\lambda)[u] =
\frac{1}{\psi(\lambda)^2}\,\D\,\psi(\lambda)^2\,\D\,\psi(\lambda)^2\,\D\,\frac{1}{\psi(\lambda)^2}\,\D^{-1}\,
.\] It is remarkable that the above operator is \emph{linear} in
$\lambda$, which is a consequence of the constraint
(\ref{eq:lin-eig}).

The kernel of the operator $R(\lambda)[u]$ is easily found to be
composed by three nonlocal vectors:
\begin{equation}\label{eq:negKdV}
\begin{array}{rcl}
u_t =     {V}\ud{}{(-1)}{1}(\lambda)[u] & = & \left({\psi(\lambda)^2}\right)_x \,,\\
u_t =     {V}\ud{}{(-2)}{1}(\lambda)[u]& = & \left(\psi(\lambda)\,\bar{\psi}(\lambda)\right)_x \,,\\
u_t ={V}\ud{}{(-3)}{1}(\lambda)[u] & =
&\left(\bar{\psi}(\lambda)^2\right)_x\,, \
\end{array}
\end{equation}
where $\psi(\lambda),\,\bar{\psi}(\lambda)$ (see equation
(\ref{eq:2ndSol})) are two l.i. solutions of the constraint
(\ref{eq:lin-eig}). These vectors contain the whole negative
hierarchies if $\lambda$ is left arbitrary, as lemma \ref{lemLeft}
states. A discrete infinite--dimensional representation of these
negative vectors, as nonlocal symmetries of the KdV equation, is
known. \cite{Gut93} On the other hand, the continuous
representation (\ref{eq:negKdV}) of these vectors, more suitable
for field theory, allows for considering them also as possible
evolution equations, which will be recognized as known integrable
equations in subsection \ref{subsec:examples neg}.

\subsection{Explicit form and Kernel of the Inverse Hereditary Operator $R^{-1}$}

The inverse hereditary operator, $R^{-1}(\lambda)$, is found
easily after inverting every factor. Assuming appropriate boundary
conditions on the field $\psi$, we get:
\[R^{-1}(\lambda)[u] =
\D\,{\psi(\lambda)^2}\,\D^{-1}\,\frac{1}{\psi(\lambda)^2}\,\D^{-1}\,\frac{1}{\psi(\lambda)^2}\,\D^{-1}\,{\psi(\lambda)^2}\,\,
.\]
 It is easy to show that the kernel of this operator is generated by the
 vector $V\ud{}{(1)}{1} = -u_x\,.$

\subsection{Factorization of Positive and Negative Symplectic and Hamiltonian Operators}

Using the factorized form of the hereditary operators, we get
easily the symplectic operators in factorized form. For the
positive ones, we have
\begin{equation*}
\begin{array}{rcl}
\Sigma^{(2)}[u] &=&
\D^{-1}\,\frac{1}{\psi^2}\,\D\,\psi^2\,\D\,\psi^2\,\D\,\frac{1}{\psi^2}\,\D^{-1}\,,
\\
\Sigma^{(3)}[u] &=&
\D^{-1}\,\frac{1}{\psi^2}\,\D\,\psi^2\,\D\,\psi^2\,\D\,\frac{1}{\psi^2}\,\D^{-1}\,\frac{1}{\psi^2}\,\D\,\psi^2\,\D\,\psi^2\,\D\,\frac{1}{\psi^2}\,\D^{-1}\,,
\end{array}
\end{equation*}
and so on, where $\psi=\psi(\lambda=0)\,.$ Notice that the
inverses of these operators give new nonlocal Hamiltonian
operators for KdV.

For the negative ones, on the other hand, we have
\begin{equation}\label{eq:neg sym}
\begin{array}{rcl}
 \Sigma^{(0)}[u] &=&
 {\psi^2}\,\D^{-1}\,\frac{1}{\psi^2}\,\D^{-1}\,\frac{1}{\psi^2}\,\D^{-1}\,{\psi^2}\,,
\\
\Sigma^{(-1)}[u] &=&
{\psi^2}\,\D^{-1}\,\frac{1}{\psi^2}\,\D^{-1}\,\frac{1}{\psi^2}\,\D^{-1}\,{\psi^2}\,\D\,{\psi^2}\,\D^{-1}\,\frac{1}{\psi^2}\,\D^{-1}\,\frac{1}{\psi^2}\,\D^{-1}\,{\psi^2}\,,
\end{array}
\end{equation}
and so on. By the way, the above expression for $\Sigma^{(0)}[u]$
turns out to solve a puzzle in the recent literature, \cite{Nut00}
for it is the inverse of Magri's Hamiltonian operator. As we see,
we have got all inverses (in factorized form) of all Hamiltonian
operators within the multi--Hamiltonian structure for KdV.

There is another, concise way to write these negative operators,
which resembles the way we wrote the negative vectors in terms of
the $\lambda$--dependent first one. We state the lemma without
proof:
\begin{lem}\label{lem:symp}
The negative symplectic operators for KdV are written in terms of
$\Sigma^{(0)}(\lambda)[u] \equiv \Sigma^{(1)}\cdot
R^{-1}(\lambda)[u]$ in the following way:
\begin{equation*}
\Sigma^{(-n)}[u] = \left.\frac{1}{{n}!}(-4)^{-{n}}
\frac{\partial^{{n}}}{\partial\lambda^{{n}}}
\Sigma^{(0)}(\lambda)[u]\right|_{\lambda=0}\,,\quad {n} \geq 1\,.
\end{equation*}
\end{lem}

A similar formula may be written for the nonlocal Hamiltonian
operators.

\subsection{Negative Vectors as Kernel of Positive Symplectic Operators}

The kernel spaces ${\mathrm{Ker}} \Sigma^{(n)}[u]\,,$ for $
{n}=1,\,\ldots,\,\infty\,,$ are easily computed in terms of the
kernel of positive powers of $R$, from the fact that
${\mathrm{Ker}} \Sigma^{(1)}[u]$ is null. We get
\begin{equation*}
{\mathrm{Ker}} \Sigma^{(n+1)} =
{\mathrm{span}}\{V\ud{}{(k)}{m}[u]\,,\quad {k} =
-1,\,-2,\,-3,\quad {m} = 1,\ldots,{n}\, \}\,,\quad {n} \geq 0 \,.
\end{equation*}

\subsection{Positive Vectors as Kernel of Negative Symplectic Operators}

Finally we compute the kernel spaces ${\mathrm{Ker}}
\Sigma^{(n)}[u]\,,$ for ${n}=0,\,-1,\,\ldots,\,-\infty$. It is
easily seen that the operator $\Sigma^{(0)}[u]$ has a null kernel.
This time we have to evaluate the kernel of negative powers of
$R$. We get
\begin{equation*}
{\mathrm{Ker}} \Sigma^{(-n)}[u] =
{\mathrm{span}}\{V\ud{}{(1)}{m}[u]\,,\quad {m} = 1,\ldots,{n}\,
\}\,,\quad {n} > 0 \,,
\end{equation*}
so that, in particular, the action principle associated to
$\Sigma^{(-2)}[u]$ for the KdV equation has the translation vector
as well as the KdV vector as generators of its kernel, therefore
the action has to be time--reparametrization invariant
\cite{Bus02} (we will see an example soon).

\subsection{Action Principles for KdV: Positive Lagrangian Ladders}

\begin{rem}If $P\d{}{a}\delta u\u{}{a}$ denotes a $1$--form,
where ${\mathrm{a}} \equiv x$ is a continuous index, we will write
the component $P\d{}{a}$ of the $1$--form as ${\mathcal{P}}(x,t)$
(which looks more like a density) when dealing with it inside an
integral sign.
\end{rem}

Following theorems \ref{thm:1} and \ref{thm:2}, we write the
action principles from equation (\ref{eq:action-KdV}):
\begin{equation}\label{eq:action ladder KdV}
S^{(m)}[u(x,t)] = \int_{t_-}^{t_+} \, \int_{x_-}^{x_+}
\,{\mathcal{P}}^{(m)}[u](x,t)\,\left(u_t + \frac{m}{m + 3/2}\,
(u_{xxx}+12\,u\,u_x))\right)\,dx\,dt\,,
\end{equation}
for ${m} > 0\,,$ where
\begin{equation*}
{\mathcal{P}}^{(1)}[u](x,t) = i_{\eta\d{}{gal}}\Sigma^{(2)}[u] =
-(\D + 4\,u\,\D^{-1} + 4\,\D^{-1}\,u)\frac{1}{8} =
-\frac{1}{2}\left(x\,u + \D^{-1}(u)\right)\,,
\end{equation*}
and successive $1$--forms are defined by recurrence:
${\mathcal{P}}^{(m+1)}[u](x,t) = R^\dag[u] \cdot
{\mathcal{P}}^{(m)}[u](x,t)$, where ${R}^\dag[u]$ is the transpose
Nijenhuis operator.

The action principle $S^{(1)}[u(x,t)]$ gives rise to the following
Euler--Lagrange equations:
\begin{equation*}
  \D^{-1}(u_t + u_{xxx}+12\,u\,u_x) = 0\,,
\end{equation*}
which are equivalent to KdV. The associated constant of the motion
is:
\begin{equation*}
  H^{(1)}[u] \equiv i_{V_{2}} P^{(1)} = \frac{1}{2}\int dx\,\left(x\,u
+ \D^{-1}(u)\right)\left(u_{xxx}+12\,u\,u_x\right) =
\frac{5}{4}\,\int dx\,\left(u_x^2-4\,u^3\right)\,,
\end{equation*}
which is a member of the known set.

The next action principle is written as equation (\ref{eq:action
ladder KdV}), with
\begin{equation*}
{\mathcal{P}}^{(2)}[u](x,t) =
R^\dag[u]\cdot{\mathcal{P}}^{(1)}[u](x,t) =
-\frac{1}{2}\left(3\,u_x + x\,u_{xx} +6\,x\,u^2 + 4\,u\,\D^{-1}(u)
+ 6\,\D^{-1}(u^2) \right)\,.
\end{equation*}
The Euler--Lagrange equations are: $2\,\left(\D + 4\,u\,\D^{-1} +
4\,\D^{-1}\,u\right)(u_t + u_{xxx}+12\,u\,u_x) = 0\, $ or, in
factorized form,
 $\,2\,\D^{-1}\,\frac{1}{\psi^2}\,\D\,\psi^2\,\D\,\psi^2\,\D\,\frac{1}{\psi^2}\,\D^{-1}\,(u_t
+ u_{xxx}+12\,u\,u_x) = 0\,.\,$ These equations are not equivalent
to the original equations: instead, they are equivalent to the
deformed equations
\begin{equation*}
  u_t =- u_{xxx} - 12\,u\,u_x + \theta_1\,\left(\psi^2\right)_x+ \theta_2\,\left(\psi\,\bar{\psi}\right)_x+ \theta_3\,\left(\bar{\psi}^2\right)_x\,,
\end{equation*}
where $\theta_j$ are arbitrary $0$--forms that multiply the
generators of ${\mathrm{Ker}} \Sigma^{(2)}[u]\,,$ and $\psi =
\psi(\lambda=0)\,.$

We could continue this process \textbf{constructively}, obtaining
explicitly the constants of the motion and the action principles
as well as the Euler--Lagrange equations: the original KdV
equation gets deformed with vectors in the negative hierarchies,
as it was mentioned before.

\subsection{Action Principles for KdV: Negative Lagrangian Ladders}
\label{subsec:neg lag KdV} Now we turn to the construction of the
negative action principles for KdV, whose action functionals are
defined by
\begin{equation*}
S^{(m)}[u(x,t)] = \int_{t_-}^{t_+} \, \int_{x_-}^{x_+}
\,{\mathcal{P}}^{(m)}[u](x,t)\,\left(u_t + \frac{m}{m + 3/2}\,
(u_{xxx}+12\,u\,u_x))\right)\,dx\,dt\,,
\end{equation*}
for ${m}<0\,.$ We only need to evaluate
${\mathcal{P}}^{(-m)}[u](x,t) = (R^\dag[u])^{-m} \cdot
{\mathcal{P}}^{(0)}[u](x,t)$, for $m<0\,.$ This is done easily
after stating the following corollary (from lemma \ref{lem:symp}):

\begin{cor}\label{cor:neg1form}
The negative $1$--forms are obtained from the first negative one
as follows:
\begin{equation*}
{\mathcal{P}}^{(-n-1)}[u](x,t) = \left.\frac{1}{{n}!}(-4)^{-{n}}
\frac{\partial^{{n}}}{\partial\lambda^{{n}}}
{\mathcal{P}}^{(-1)}(\lambda)[u](x,t)\right|_{\lambda=0}\,,\quad
{n} \geq 1\,,
\end{equation*}
where ${\mathcal{P}}^{(-1)}(\lambda)[u](x,t) =
-\frac{1}{8}\,\psi^2\,\D^{-1}\,\frac{1}{\psi^2}\,\D^{-1}\,\frac{1}{\psi^2}\,\D^{-1}\psi^2$
or, using equation (\ref{eq:local--nonlocal}),
\begin{equation*}
{\mathcal{P}}^{(-1)}(\lambda)[u](x,t) =
\frac{1}{16}\,\left(\psi_\lambda
\,\bar{\psi}-\psi\,\bar{\psi}_\lambda \right)\,.
\end{equation*}
\end{cor}
The first negative action principle from equation (\ref{eq:action
ladder KdV}) is thus:
\begin{equation*}
S^{(-1)}[u(x,t)] = \int_{t_-}^{t_+} \, \int_{x_-}^{x_+}
\,\frac{1}{16}\,\left( \psi_\lambda
\,\bar{\psi}-\psi\,\bar{\psi}_\lambda \right)\,\left(u_t +
\frac{-1}{-1 + 3/2}\, (u_{xxx}+12\,u\,u_x))\right)\,dx\,dt\,,
\end{equation*}
or, after some manipulations, $S^{(-1)}[u(x,t)] = \int_{t_-}^{t_+}
\, \int_{x_-}^{x_+} \,\left(\frac{1}{16}\,\left(\psi_\lambda
\,\bar{\psi} -\psi\,\bar{\psi}_\lambda \right)\,u_t
-\frac{1}{4}\,u\right)\,dx\,dt\,, $ where we have to evaluate the
fields $\psi,\,\bar{\psi}$ at $\lambda = 0\,.$ This action
principle is highly nonlocal, even in terms of the auxiliary
fields (see equation (\ref{eq:local--nonlocal})). However, the
Euler--Lagrange equations are obtained as usual, varying the
action with respect to the field $u$, and using the appropriate
transformation matrices. We obtain $-\Sigma^{(-1)}[u]\cdot(u_t +
u_{xxx}+ 12\,u\,u_x) = 0\,,\,$ or explicitly, using the fact that
the kernel of this operator is generated by $V\ud{}{(1)}{1}\,,$
\begin{equation*}
  u_t =- u_{xxx}- 12\,u\,u_x + \theta_1 \,u_x\,,
\end{equation*}
where $\theta_1$ is arbitrary.

We stress there is no need to hesitate about the inclusion of
auxiliary fields in the negative action principles, for they are
not varied independently. Alternatively, we may map the above
action into a mixed action principle, in which the fields $\psi$,
$u$ and a Lagrange multiplier $\rho$ are varied independently:
\begin{equation}\label{eq:action -1 Lax}
S^{(-1)}[\psi(x,t),\, u(x,t),\, \rho(x,t)] = \int_{t_-}^{t_+} \,
\int_{x_-}^{x_+}
\,\left(\frac{1}{16}\,\frac{\psi_t}{\psi^3}\left(\D^{-1}
\psi^2\right) + \frac{1}{8}\,\frac{\psi_{xx}}{\psi} + \rho \left(u
+ \frac{\psi_{xx}}{2\,\psi}\right) \right)\,dx\,dt\,.
\end{equation}
See \cite{Nut01} for a general discussion.

The next negative action principles are quite simple. Recall that
the evolution vector itself $V_{2}$ is in the kernel of the
symplectic operators $\Sigma^{(m)}\,,$ for ${m} = -2,\ldots,
-\infty\,,$ so that the action principles should be
time--reparametrization invariant. \cite{Bus02} From equation
(\ref{eq:action-KdV}), the only chance is $K^{(m)}\propto
i_{V_{2}}P^{(m)} = 0$ or a numeric constant (which would not
change the action principle), for ${m}\leq -2\,.$ This is easily
shown, from the fact that the interior product $I(\lambda) =
i_{V\ud{}{(1)}{1}}P^{(-1)}(\lambda)[u] =
-\frac{1}{16}\,\int_{x_-}^{x_+} dx\,\left(\psi_\lambda
\,\bar{\psi} -\psi\,\bar{\psi}_\lambda \right)\,u_x\,, $ where
$\psi=\psi(\lambda)\,,$ is a numeric constant for all $\lambda$:
$I(\lambda) = -\frac{1}{32}(x_+-x_-)\,.$ We get, then, manifestly
time--reparametrization invariant actions:
\begin{equation*}
S^{(-n-1)}[u(x,t)] =
\frac{1}{16\,({n}!)}\,(-4)^{-{n}}\,\int_{t_-}^{t_+} \,
\int_{x_-}^{x_+}
\,\frac{\partial^{{n}}}{\partial\lambda^{{n}}}\left.\left(
\psi_\lambda \,\bar{\psi}-\psi\,\bar{\psi}_\lambda
\right)\right|_{\lambda=0}\,u_t\,dx\,dt\,,\quad {n} \geq 1\,,
\end{equation*}
and we may use equation (\ref{eq:local--nonlocal}) in order to
write the $\lambda$--derivatives in terms of nonlocal expressions.
For example, the second negative action principle is
\begin{equation*}
S^{(-2)}[u(x,t)] = -\frac{1}{64}\,\int_{t_-}^{t_+} \,
\int_{x_-}^{x_+} \,\left.\left( \psi_{\lambda \lambda}
\,\bar{\psi}-\psi\,\bar{\psi}_{\lambda \lambda}
\right)\right|_{\lambda=0}\,u_t\,dx\,dt\,,
\end{equation*}
where $\psi_{\lambda\lambda} = 2
\,{\psi}\,\D^{-1}\,\frac{1}{\psi^2}\,\D^{-1}\,{\psi}^2\,\D^{-1}\,\frac{1}{\psi^2}\,\D^{-1}\,{\psi}^2$
and $\bar{\psi} = \psi\,\D^{-1}(1/\psi^2)\,;$ the Euler--Lagrange
equations are equivalent to
\begin{equation*}
u_t = \theta_1\,\left(u_{xxx} + 12\,u\,u_x\right) +
\theta_2\,u_x\,
\end{equation*}
where, as usual, $\theta_j$ are arbitrary functionals. The
invariance $t \to \tau(t)$ is evident.

\subsection{The Missing Action Principle for KdV, a time--dependent Constant of the Motion and the Internal Vectors}

So far we have obtained two ladders of action principles for the
KdV equation: the positive (quasi local) and the negative ladders
(highly nonlocal). However, there is a missing action principle:
this is the case ${m} = 0$, which is actually twofold: first, the
$1$--form $P^{(0)}[u] \equiv i_{\eta\d{}{gal}}\Sigma^{(1)}[u] =
-x/8 $ is closed: $P^{(0)}[u] = \delta C^{(0)}[u]\,,$ where
$C^{(0)}[u] = -\int_{x_-}^{x_+} dx\,x\,u/8\,.$ From equation
(\ref{eq:t--d const}), we obtain a known \cite{Eil81}
time--dependent constant of the motion for KdV:
$C[u,t]=\frac{1}{8}\,\int_{x_-}^{x_+} (6\,t\,u^2-x\,u)\,dx\,.$

Second, the action principle for the symplectic $2$--form
$\Sigma^{(0)}[u]$ (see equation (\ref{eq:neg sym})) has to be
evaluated by hand. After some hard but straightforward
calculations, we find that the $1$--form
$\,{\mathcal{P}}^{(M)}[u](x,t) \equiv \frac{\psi^2}{4}
\D^{-1}\left(\frac{1}{\psi^2}\,\ln \psi\right)\,$ is a solution of
$\,\delta {P}^{(M)} = \Sigma^{(0)}\,.$

In this case, we map to the $\psi$--coordinate system for
simplicity. We get the action principle
\begin{equation*}
S^{(0)}[\psi,u,\rho] =
-\frac{1}{8}\,\int_{x_-}^{x_+}\left(\frac{\psi_x\,\psi_t}{\psi^2}
- \frac{\psi_{xx}^2}{\psi^2} + \rho\,\left(u +
\frac{\psi_{xx}}{2\,\psi}\right)\right)\,dx\,dt\,.
\end{equation*}
The Euler--Lagrange equations we obtain are: for the field $\psi$,
\begin{equation*}
\frac{1}{\psi}\,\D\,\frac{1}{\psi} \left(\psi_t + \psi_{xxx} -
3\frac{\psi_x\psi_{xx}}{\psi}\right) = 0\,,
\end{equation*}
for the field $u$: $u =-\frac{\psi_{xx}}{2\,\psi}
\,\Rightarrow\,u_t =
  -u_{xxx}-12\,u\,u_x\,,$
and for the Lagrange multiplier:  $\rho = 0\,.$

Notice that the Euler--Lagrange equations for $\psi$ are
equivalent to
\begin{equation*}
  \psi_t =- \psi_{xxx} +
3\frac{\psi_x\psi_{xx}}{\psi} + \theta_1\,\psi\,,
\end{equation*}
where $\theta_1$ is arbitrary. This symmetry is one of the three
known \cite{Gut93} internal symmetries (i.e., those which do not
affect the field $u$) of the eigenvalue problem
(\ref{eq:lin-eig}). In the $\psi$--coordinate system we write it
as $\psi_t =  V\ud{}{(-2)}{0}[\psi] = -\frac{1}{2}\,\psi\,$ (the
numeric factor is only for simplicity). In the cited reference it
is shown that all negative vectors and the internal symmetries
span a loop algebra over $SL(2,{\mathbb{R}})$.

\subsection{Nonlocal Constants of the Motion for KdV}
\label{subsec:nonlocal-const-KdV-2} As a final result, we will
construct explicitly three new sets of constants of the motion for
the KdV equation, starting from the nonlocal objects we have
obtained. We denote $\psi=\psi(x,t;\lambda=0)$ for simplicity from
here on, unless explicitly stated. Consider the action principle
(\ref{eq:action -1 Lax}): the term which multiplies the velocity
$\psi_t$ is the mapping of the Lagrangian $1$--form $P^{(-1)}[u]$
into $\psi$--coordinates:
\begin{equation*}
{\mathcal{P}}^{(-1)}[\psi](x,t) =
\frac{1}{16\,\psi^3}\left(\D^{-1} \psi^2\right)\,.
\end{equation*}
On the other hand, from theorem \ref{thm:2} we have
$\Lie{V_2}{}{P}^{(-1)} = \delta K^{(-1)}\,,$ where $K^{(-1)}
=\frac{3}{16}\,\int_{x_-}^{x_+}dx\,\frac{\psi_x^2}{\psi^2}\,.$
Consider now the $0$--form
\begin{equation*}
  H^{(-2)}[\psi] \equiv -16\,i_{V\ud{}{(-2)}{0}}P^{(-1)} =
 - \int_{x_-}^{x_+}dx\,\frac{1}{\psi^2}\,\left(\D^{-1}\,\psi^2\right)\,
\end{equation*}
or, more concisely, $H^{(-2)}[\psi] =
\int_{x_-}^{x_+}dx\,\psi\,\bar{\psi}\,. $

We use Leibnitz rule to show that this is a constant of motion for
the KdV equation, which in $\psi$--coordinates reads $V_2[\psi] =
- \psi_{xxx} + 3\frac{\psi_x\psi_{xx}}{\psi}\,:$ as
$\Lie{V_2}{V\ud{}{(-2)}{0}}[\psi] = 0\,,$ we find:
$\Lie{V_2}{}H^{(-2)}[\psi] = \Lie{V\ud{}{(-2)}{0}}{K^{(-1)}}[\psi]
=
  0\,,$ where the last equality comes from the fact that $K^{(-1)}[\psi]$ is
invariant under scaling of $\psi$.

Thus $H^{(-2)}[\psi]$ is a nonlocal constant of the motion for the
KdV flow. But if we recall that the fields $\psi,\,\bar{\psi}$ are
solutions of the linear problem (\ref{eq:lin-eig}), and that $u$
does not change if these fields are replaced by other arbitrary
linear combinations, we get indeed three constants of the motion:
$H^{(-1)}[\psi]  = \int_{x_-}^{x_+}dx\,\psi^2\,,\, $
 $H^{(-2)}[\psi]  = \int_{x_-}^{x_+}dx\,\psi\,\bar{\psi}\,,\,$
 $H^{(-3)}[\psi]  = \int_{x_-}^{x_+}dx\,\bar{\psi}^2\,. $

There is a reference that supports this construction: in
\cite{Lax68}, in the context of the eigenvalue ``Schr\"odinger"
problem (\ref{eq:lin-eig}), the author assumes that the total
probability (here denoted by $H^{(-1)}[\psi]$) is equal to $1$.
But it is indeed a constant of motion of its own! Moreover, these
are indeed special cases ($\lambda=0$) of more general constants
of motion. Along the same lines, we get three families,
parametrized by the eigenvalue $\lambda$:
\begin{equation}\label{eq:nonlocal constants}
 \begin{array}{rcl}
    H^{(-1)}(\lambda)[\psi] & = & \int_{x_-}^{x_+}dx\,\psi(\lambda)^2\,, \\
    H^{(-2)}(\lambda)[\psi] & =& \int_{x_-}^{x_+}dx\,\psi(\lambda)\,\bar{\psi}(\lambda)\,, \\
    H^{(-3)}(\lambda)[\psi] & =& \int_{x_-}^{x_+}dx\,\bar{\psi}(\lambda)^2\,. \
  \end{array}
\end{equation}
These are real new constants (indeed they contain, in their Taylor
series around $\lambda=0$, the constants of the motion from
theorem \ref{thm:3}). In order to evaluate them explicitly, take,
for example, successive derivatives of the first one with respect
to $\lambda$, evaluate at $\lambda=0$ and use equation
(\ref{eq:local--nonlocal}). We get:
\begin{equation*}
Q^{(-1;n)}[\psi] \equiv
\int_{x_-}^{x_+}dx\,\psi\,L^{-n}\,\psi\,,\quad n=
0,\ldots,\,\infty\,,
\end{equation*}
and we see they are increasingly nonlocal constants of the motion.

It is worth to mention that these nonlocal constants, when mapped
to the coordinate system in which the KdV equation maps into the
Harry--Dym equation (see \cite{Gol91}), reproduce the results
obtained independently in a recent work \cite{Bru02}, and add
three more constants to the Harry--Dym equation: the mappings of
the nonlocal constants of motion (\ref{eq:nonlocal constants}) for
$\lambda=0\,$ into the Harry--Dym equation $\omega_t =
(\omega^{-1/2})_{sss}$ for the field $\omega(s,t)\,,$ are
$H^{(-1)}[\omega]
 =  \int_{s_-}^{s_+}ds\,\omega\,,$
 $H^{(-2)}[\omega]  =
\int_{s_-}^{s_+}ds\,s\,\omega\,,$
 $ H^{(-3)}[\omega]  =
\int_{s_-}^{s_+}ds\,s^2\,\omega\,. $

\section{Results for other KdV Positive and Negative equations}
\label{sec:results negative}
\subsection{Some Positive, Negative and Internal Vectors as known
Integrable Equations} \label{subsec:examples neg} We write the
internal vectors after transformation to Schwartzian coordinates,
defined by $ \zeta_x(x,t;\lambda) =  {\psi(x,t;\lambda)^{-2}}\,. $
The internal vectors in $\zeta$--coordinates are
$V\ud{}{(-1)}{0}[\zeta] =  1\,,$
 $V\ud{}{(-2)}{0}[\zeta]  =  \zeta\,,$
 $V\ud{}{(-3)}{0}[\zeta]  = \zeta^2\,.$

The vectors $V\ud{}{(-3)}{0}\,,\,(V\ud{}{(-3)}{0} -
V\ud{}{(-1)}{1})/2\,,$ and $V_2$ give the evolution equations $
\zeta_t =  \zeta^2\,,$ $\zeta_t = \frac{1}{2}\,\zeta^2 -
\frac{1}{8}\frac{\zeta_{x
  \lambda}}{\zeta_x}\,,$ and $\zeta_t = 6\,\lambda\,\zeta_x +
   \frac{3\,{\zeta_{xx}}^2}{2\,
      \zeta_x} -
   \zeta_{xxx}\,,$ where the fields are evaluated at $\lambda=0\,.$ The last of these
equations is a special case of the Krichever--Novikov equation,
\cite{Mal01a} and the first and the second equations, via the
transformation $z = \ln\left(2\,\zeta_x\right)\,,$ may be mapped
to the Liouville equation $z_{x t} = \exp {z}\,,$ and the ShG
equation $z_{x t} = \sinh z\,.$ For completeness, we just mention
that the associated Camassa--Holm equation \cite{Cam93} and the
Hunter--Zheng equation \cite{Bru02} are obtainable from the
negative vectors $V\ud{}{(-2)}{1}$ and $V\ud{}{(-1)}{1}$,
respectively, via suitable coordinate transformations, and that
the Harry--Dym equation, \cite{Gol91} just like the above case of
the Krichever--Novikov equation, is a mapping of the KdV equation.

\subsection{Action Principles for the Sinh--Gordon Equation}
\label{subsec:action-sinhg}As a representative of the extension of
the results on action principles for equations in the negative
hierarchies, we work out some examples for the ShG equation. The
results in this subsection are new up to our knowledge, except
when it is explicitly stated. We will work in the $z$--coordinate
system, where the ShG equation is $z_t = \d{V}{ShG}[z] =
\D^{-1}\,\sinh z\,.$

\subsubsection{Pure ShG equation: Symplectic matrix $\Sigma^{(0)}$\,}

We look for a standard Lagrangian pair for the ShG equation of the
form (${P}^{(M)}[z]$;\,${K}^{(M;ShG)}[z]$), where
${\mathcal{P}}^{(M)}[z] = -\frac{1}{32}  z_x\,$ is the mapping of
${\mathcal{P}}^{(M)}[u]$ to $z$--coordinates. The symplectic
$2$-form $\u{\Sigma}{(0)}[z] = \frac{1}{16}\, \D $ has only one
vector in the kernel, namely ${V}\ud{}{(-2)}{0}[z] = 4\,.$ On the
other hand, the standard Lagrangian $0$--form solves $\nonumber
\frac{\delta}{\delta z} {K}^{(M;ShG)}[z]  =
\Lie{\d{V}{ShG}}{{P}^{(M)}[z]}= -\frac{1}{32} \left(\sinh z -
z\,\cosh z \right) \,.$ We get after integration the usual action
principle for the ShG equation: \cite{Eil81}
\[S[z(x,t)] = \frac{1}{32}\int_{t_-}^{t_+} dt\,dx\,\left(-z_x\,z_t - 2\,\cosh z\right)\,,\]
and the Euler--Lagrange equations are simply $z_t  =
\D^{-1}\,\sinh z + \theta_1 \,,\,$ where $\theta_1$ is arbitrary.

\subsubsection{ShG equation deformed with first Positive Vector: Symplectic
matrix $\Sigma^{(-1)}$\,}

The next negative $1$--form, $\u{P}{(-1)}$, reads
$\u{\mathcal{P}}{(-1)}[z](x,t) = -\frac{1}{32} \,\,
{\mathrm{e}}^z\,\left(\D^{-1}\,{\mathrm{e}}^{-z}\right)\,.$ The
associated symplectic $2$--form is $\u{\Sigma}{(-1)}[z] =
-\frac{1}{32} \left({\mathrm{e}}^{z} \,\D^{-1}\,{\mathrm{e}}^{-z}
+ {\mathrm{e}}^{-z} \,\D^{-1}\,{\mathrm{e}}^{z}\right)\,,$ which
inherits the kernel (generated from $\ud{V}{(1)}{1}[z] = - z_x $)
from that in the $u$--coordinate system only for special boundary
conditions: defining the boundary terms $\overline{f} \equiv f_+
+f_-,$ and $f_\pm = f(x_\pm),$ the expression
\[\u{\Sigma}{(-1)}[z] \cdot \ud{V}{(1)}{1}[z] = \frac{1}{64} \, \left({\mathrm{e}}^z \,\overline{{\mathrm{e}}^{-z}} - {\mathrm{e}}^{-z} \,\overline{{\mathrm{e}}^{z}}\right)\,\]
is zero only for boundary conditions $z_+ = z_- + i \pi (2\,{{n}}
+ 1)\,,\quad {{n}} \in {\mathbb{Z}}$.

For other boundary conditions, however, this Lagrange bracket has
no kernel, which will show up in the variational principle for the
ShG vector by the fact that the Euler--Lagrange equations get
deformed by a factor of the vector $\ud{V}{(1)}{1}[z]$, which is
not arbitrary: it depends on the boundary conditions used for the
$z$--coordinates.

In the generic case when $\overline{{\mathrm{e}}^{z}}\neq 0$
(invertible symplectic $2$--form $\Sigma^{(-1)}$), the action
principle is explicitly
\[S[z(x,t)] = \int_{t_-}^{t_+} dt\,\left[-\frac{1}{32}\int_{x_-}^{x_+} dx\, {\mathrm{e}}^z(\D^{-1}{\mathrm{e}}^{-z})(z_t - \D^{-1}\sinh z) + \u{K}{(-1;ShG)}[z]  \right]\,,\]
 and the Euler--Lagrange equations are
\[\frac{1}{32}\left({\mathrm{e}}^{z} \,\D^{-1}\,{\mathrm{e}}^{-z} + {\mathrm{e}}^{-z} \,\D^{-1}\,{\mathrm{e}}^{z}\right)\left(z_t - \D^{-1} \sinh z + \theta[A_+,\,A_-]\,z_x\right) = 0\,\]
or, equivalently,
\[z_t - \D^{-1} \sinh z + \theta[A_+,\,A_-]\,z_x = 0\,,\]
where $\u{K}{(-1;ShG)}[z]  = \int_{x_-}^{x_+} dx\,{\mathrm{e}}^z
(\D^{-1}{\mathrm{e}}^{-z})^2/128 + F[A_+,\,A_-]\,,$ $\quad A_{\pm}
\equiv \int_{x_-}^{x_+}dx\,{\mathrm{e}}^{\pm z}\,,$ and $\theta
,\,F$ solve the equation:
\begin{equation}\label{Fa}
\delta F[A_+,\,A_-]  =
-\theta[A_+,\,A_-]\,(\overline{{\mathrm{e}}^{-z}}\,\delta A_+ +
\overline{{\mathrm{e}}^{z}}\,\delta A_-) + \frac{1}{8} \,A_-\,(A_-
- 2\, A_+)\,\delta A_+\,.
\end{equation}

There are many solutions of the above equation for a given set of
boundary conditions on the limiting values of $z_\pm$, so we
discuss, as examples, only two representative, nonintersecting
cases of boundary conditions, for which the symplectic $2$--form
$\Sigma^{(-1)}$ is invertible:

\begin{enumerate}
\item[(i)]{$z_+ = -z_- + i \pi (2\,{{n}})\,,\quad {{n}} \in
{\mathbb{Z}}\,; \quad \cosh z_+ \neq 0\,.$}

A solution of equation (\ref{Fa}) is $\theta = - A_+\,A_- /{4
\,\overline{{\mathrm{e}}^z}}\,,\quad F= A_+\,A_-^2/512 \,,$
 which is well defined because of the boundary conditions
used.

\item[(ii)]{$z_+ = -z_- + i \pi (2\,{{n}} + 1)\,,\quad {{n}} \in
{\mathbb{Z}}\,; \quad \sinh z_+ \neq 0\,.$}

In this case, a solution of equation (\ref{Fa}) is
\[\theta = - \left(A_+^2 - A_-^2 \right)/{8
\,\overline{{\mathrm{e}}^z}}\,,\qquad F= - \left(A_+^2\,A_- -
\frac{1}{3} (A_+^3 + A_-^3) \right)/{512}\,.\]

The usual constant of the motion for the ShG equation, $H[z] =
\int_{x_-}^{x_+} dx\,\cosh z\,,$ works in this case also: under
the boundary conditions used, we get
\[\dot{H}[z] = \int_{x_-}^{x_+} dx\,\sinh z \,(\D^{-1}\, \sinh z - \theta\, z_x)
= \theta \,\overline{\cosh z} = 0\,.\]
\end{enumerate}

\subsection{Alternative Lax Pairs and Constants of the Motion for Negative
Equations} In reference \cite{Bru95}, the authors find alternative
Lax pairs for the KdV equation (as well as for every evolution
equation in the KdV positive hierarchy) by making no ansatz: they
just use the evolution equation and the hereditary operator.

We present a similar construction, this time for the negative
vectors. By so doing we are answering an open question in
reference \cite{Bru02}.

As it is shown in \cite{Bru95}, given an evolution equation $u_t =
V[u]$, and a recursion operator $R$ for $V$ (i.e., $\Lie{V}{\,R}=
0\,$), it follows that
\begin{equation}\label{eq:Lax pair}
  \frac{\D}{\D t}R = \left[V',\,R\right]\,,
\end{equation}
where $V'$ denotes the Frechet derivative, and the square bracket
is the commutator. The above equation defines the alternative Lax
Pair $(R,\,V')$.

Now, take $R$ as the hereditary operator $R[u]$, and $V$ as the
negative vector $V\ud{}{(-1)}{1}(\lambda)[u] =
(\psi(\lambda)^2)_x$, for arbitrary $\nu,\,\lambda\,.$ We need to
evaluate the Frechet derivative of this vector with respect to the
field $u$. Using the transformation matrix $\frac{\delta
\psi}{\delta u} =
-2\,\psi\,\D^{-1}\,\frac{1}{\psi^2}\D^{-1}\,\psi^2\,,$ where
$\psi$ stands for $\psi(\lambda)$ from here on, we get
$V\ud{}{(-1)}{1}(\lambda)'[u] =
-4\,\D\,\psi^2\,\D^{-1}\,\frac{1}{\psi^2}\D^{-1}\,\psi^2\,.$
 On the other hand, the hereditary operator is
\begin{equation*}
R[u] = R(\lambda)[u] - 4\,\lambda\,{\mathbb{I}}=
\frac{1}{\psi^2}\,\D\,\psi^2\,\D\,\psi^2\,\D\,\frac{1}{\psi^2}\,\D^{-1}
- 4\,\lambda\,{\mathbb{I}}\,,
\end{equation*}
and it is also written as $R[u] = \D^{2} + 8\,u +
4\,u_x\,\D^{-1}\,.$ Now we apply the Lax pair equation
(\ref{eq:Lax pair}), getting after some rearrangements the
operator equation $8\,u_t + 4\,u_{x t}\,\D^{-1} = 8\,(\psi^2)_x +
4\,(\psi^2)_{xx}\,\D^{-1}\,,$ which implies $u_t =
(\psi(\lambda)^2)_x\,.$ Recall this equation contains all the
negative vectors in the corresponding negative hierarchy, so that
the Lax pair we have presented indeed works for all vectors in
that hierarchy. Similarly, for the other two negative hierarchies
we get the Lax pairs $(R,\,B_2)$ and $(R,\,B_3)$, with
\begin{equation*}
\begin{array}{rcl}
B_2 = V\ud{}{(-2)}{1}(\lambda)'[u]& = & -4\,\D\,\psi^2\,\D^{-1}\,{\bar{\psi}}\,{\psi}^{-1}\,\D\,{\bar{\psi}}\,{\psi}^{-1}\,\D^{-1}\,\left(\bar{{\psi}}\right)^{-2}\,\D^{-1}\,\psi\,\bar{\psi}\,, \\
B_3 = V\ud{}{(-3)}{1}(\lambda)'[u]& = &
-4\,\D\,\bar{\psi}^2\,\D^{-1}\,{\left(\bar{\psi}\right)^{-2}}\,\D^{-1}\,\bar{\psi}^2\,.\
\end{array}
\end{equation*}

In this way, we may construct an infinite number of constants of
the motion for the negative vectors, from Adler traces of
positive, semi--integer powers of the Nijenhuis operator $R$:
these are just the usual (local) constants of the motion for the
KdV equation. \cite{Bru95} A natural conjecture is that Adler
traces of positive, semi--integer powers of $R^{-1}$ will give our
nonlocal constants for KdV defined in equation (\ref{eq:nonlocal
constants}). If that is true, we could infer that the nonlocal
constants of the motion for KdV should also work for the negative
vectors, which can be explicitly checked. We present the results
only for the hierarchy $V\ud{}{(-1)}{1}(\lambda)\,,$ because Lie
derivatives of the results along the internal vector
$V\ud{}{(-3)}{0}$ map the objects into similar ones for the other
two negative hierarchies:

\begin{enumerate}

\item[(i)]{}Conserved currents: defining the boundary term
$\overbrace{f} \equiv f(x_+)-f(x_-)\,,$ the integral
$H^{(-1)}(\nu)[\psi]  = \int_{x_-}^{x_+}dx\,\psi(\nu)^2\, $ solves
\begin{equation*}
\Lie{V\ud{}{(-1)}{1}(\lambda)}{H^{(-1)}(\nu)[\psi]} = -
  \overbrace{E(\nu, \lambda)[\psi]}\,,
\end{equation*}
where $E(\nu, \lambda)[\psi] \equiv
\frac{\psi(\lambda)\,\psi_x(\nu)-\psi(\nu)\,\psi_x(\lambda)}{\lambda-\nu}\,.$

\item[(ii)]{}Constants of the motion: the expression
\begin{equation*}
G^{(-1)}(\nu, \lambda)[\psi,t] = \int_{x_-}^{x_+}dx\,\psi(\nu)^2 +
t\,\overbrace{\frac{E(\nu,\lambda)[\psi]^2}{1-t\,E(\lambda,\lambda)[\psi]^2}}\,
\end{equation*}
is a constant of the motion for the flow
$V\ud{}{(-1)}{1}(\lambda)\,.$

These constants and currents are infinite in number and work for
every vector in the respective negative hierarchy, because
$\lambda$ and $\nu$ are arbitrary. By considering their Taylor
expansion around $\lambda = \nu =0,$ the explicit expression for
the constants and conserved currents for each negative vector is
easily worked out using equations (\ref{eq:local--nonlocal}) and
(\ref{eq:neg(-n) from neg(-1)}).
\end{enumerate}

\section{Conclusion}
\label{sec:conc}The Lagrangian point of view determines a unifying
scheme for the study of integrable equations belonging to
hierarchies related to hereditary operators. For all evolution
vectors in these hierarchies, nonlocal symmetries, Lax pairs,
constants of the motion, conserved currents and an infinite ladder
of action principles all come out in a constructive, explicit way
from the same structure. Moreover, new equations, which are mixed
or deformed versions of known integrable equations, arise as the
Euler--Lagrange equations of the action principles obtained. As an
example, we apply this scheme to the KdV equation, and the results
are directly mappable to other related equations in the positive
KdV hierarchies (e.g., Harry--Dym and a special case of
Krichever--Novikov equations) as well as in the negative KdV
hierarchies (e.g., Sinh--Gordon, Liouville, Camassa--Holm and
Hunter--Zheng equations): in particular, we obtain a new nonlocal
action principle for the Sinh--Gordon equation which leads to a
deformed version of this equation, and an infinite number of
nonequivalent, nonlocal action principles for KdV, possessing
time--reparametrization invariance, are explicitly found. The
construction of alternative Lax pairs for negative equations
arises naturally, without any ansatz, from this scheme, and it is
shown that negative equations essentially share the constants of
the motion (local as well as nonlocal) for the KdV equation.

\section{Acknowledgements}
M.B. wishes to mention that this work was done in part during a
research stay at the Relativity Center, The University of Austin,
Texas, U.S.A., where M.B. was invited by Prof. L. C. Shepley; M.B.
is deeply grateful to him, the professors and students in the
Center for their support and the following people whose comments
and suggestions were fundamental for his understanding of the
topics involved in the present work: P. J. Morrison, C.
deWitt--Morette, C. Valls, J. Zanelli, C. Teitelboim, R. Troncoso.

Finally, M.B. acknowledges the financial support from a
Fundaci\'on Andes Grant for Doctoral Studies, and a Conicyt Grant
for Thesis Completion.

\bibliographystyle{unsrt}
\bibliography{finalbib-kdv}

\end{document}